\documentclass[notoc,paper,letterpaper]{JHEP3}
\usepackage{amssymb}
\usepackage{epsfig}
\newcommand{\be}{\begin{equation}}
\newcommand{\ee}{\end{equation}}
\newcommand{\bea}{\begin{eqnarray}}
\newcommand{\eea}{\end{eqnarray}}
\newcommand{\ie}{{\it i.e.}}
\newcommand{\eg}{{\it e.g.}}

\newcommand{\R}{\mathbb{R}}
\newcommand{\Z}{\mathbb{Z}}
\newcommand{\C}{\mathbb{C}}
\newcommand{\U}{\mathop{\rm U}}
\newcommand{\SU}{\mathop{\rm SU}}

\newcommand{\Sp}{\mathop{\rm Sp}}
\newcommand{\bo}{\hbox{1\kern-.23em{\rm l}}}
\newcommand{\til}{\widetilde}
\def\del{{\partial}}

\def\bar{\overline}

\def\a{\alpha}
\def\b{\beta}
\def\ga{\gamma}
\def\de{\delta}
\def\la{\lambda}
\def\ad{{\dot\a}}
\def\bd{{\dot\b}}
\def\gd{{\dot\ga}}

\def\si{\sigma}
\def\th{\theta}
\def\thb{{\bar\th}}
\def\thp{\th^+}
\def\thm{\th^-}
\def\thbp{\thb^+}
\def\thbm{\thb^-}

\def\Wb{{\bar W}}

\def\Db{{\bar D}}
\def\Dp{D^+}
\def\Dm{D^-}
\def\Dbp{\Db^+}
\def\Dbm{\Db^-}
\def\Dpp{{D^{++}}}
\def\Dmm{{D^{--}}}
\def\Vpp{{V^{++}}}
\def\Vmm{{V^{--}}}

\def\dsl{{\not\!\del}}

\title{On Superspace Chern-Simons-like Terms}

\author{Philip C. Argyres$^{1,2}$, Adel M. Awad$^3$, Gregory A. Braun$^1$, and F. Paul Esposito$^1$\\
$^1$\,Department of Physics, University of Cincinnati, Cincinnati OH 45221-001, U.S.A. \\
\ \ \email{argyres,braun,esposito@physics.uc.edu}
\vskip5pt
$^2$\,School of Natural Sciences, Institute for Advanced Study, Princeton, NJ 08540
\vskip5pt
$^3$\,Department of Physics, Faculty of Science, Ain Shams University,
Cairo  11566, Egypt\\
\ \ \email{a\_awad@asunet.shams.edu.eg}}

\abstract{We search for superspace Chern-Simons-like higher-derivative terms
in the low energy effective actions of supersymmetric theories in
four dimensions.  Superspace Chern-Simons-like terms are those gauge-invariant
terms which cannot be written solely in terms of field strength superfields and
covariant derivatives, but in which a gauge potential superfield appears explicitly.
We find one class of such four-derivative terms with $N=2$ supersymmetry which, 
though locally on the Coulomb branch can be written solely in terms of field strengths,
globally cannot be.   These terms are classified by certain Dolbeault cohomology 
classes on the moduli space.  We include a discussion of other examples of 
terms in the effective action involving global obstructions on the Coulomb branch.
}

\begin{document}

\section{Introduction}

Current fundamental non-gravitational theories of nature are effective
field theories---local,  Lorentz-invariant, low energy
approximations to some complete theory.  Effective theories are
organized in a derivative expansion, where terms in the action
with fewer derivatives dominate the long wavelength, low
energy behavior.  This expansion is organized by assigning
spacetime derivatives weight $+1$ and fields various other weights
(which we will discuss below).  One then considers all terms of a
given weight that can appear in the action consistent with any
gauge invariances as well as global symmetries.  The lowest-weight
terms are the most important at low energies. This expansion is
useful when there are only a finite number
of terms of a given weight.  If a field should have negative
weight, then the derivative expansion breaks down.  In this
paper we will show that the derivative expansion in
four-dimensional theories with extended supersymmetry suffers from
this problem: vector potential multiplets have non-positive
weight.  But the way vector potentials enter into the action is
constrained by gauge invariance, so there may be, in fact, only a
finite number of gauge-invariant terms of a given weight.  The
problem that this paper faces is how to list all gauge-invariant
terms of a given weight if the gauge potential does not have
positive weight.

Let us illustrate this problem in a simple, non-supersymmetric, context.
Consider a theory of a single abelian vector field, $A_\mu$, in four dimensions.
Normally, $A_\mu$ can be assigned a positive weight, \eg, $w(A_\mu)=+1$, the
same as its scaling dimension.  In this case there are a finite number of terms
of a given weight, even without using gauge invariance.  But in theories with
extended supersymmetry, we will see that we must assign weight $w(A_\mu)=0$.
Then there are an infinite number of local, Lorentz-invariant terms for a given weight,
before imposing gauge invariance.  A commonly-held belief is that there are
no Chern-Simons-like terms in even dimensions:
all gauge invariants made just of Abelian gauge fields can be written solely in
terms of field strengths, $F_{\mu\nu}=\del_\mu A_\nu - \del_\nu A_\mu$, and
their derivatives.  If this is true, then, since $F_{\mu\nu}$ has weight $+1$,
there will be only a finite number of gauge-invariant terms of a given weight.
However, we are unaware of a proof of the absence of Chern-Simons-like terms
in even dimensions.  To prove it one should show that
for every gauge-invariant $f$ there exists a $g$ such that $\int\!d^{2n}x\,
f(A_\mu,\del_\nu) =  \int\!d^{2n}x\, g(F_{\mu\nu},\del_\rho)$ modulo
surface terms.  This is difficult to prove because the number of
ways that a $g(F_{\mu\nu},\del_\rho)$ can be written using integration
by parts as some not obviously gauge-invariant collection of terms
$f(A_\mu,\del_\nu)$ grows at least exponentially with the number of
$F$'s in $g$.  

This example gives a flavor of the type of problem that we will face in 
supersymmetric effective actions.  We emphasize, though, that in 
the supersymmetric context, the existence of \emph{superfield} 
Chern-Simons-like terms does \emph{not} necessarily imply the
existence of the hypothetical non-supersymmetric Chern-Simons-like 
terms discussed above.  Indeed, examples of superspace Chern-Simons-like 
terms which do not are known, as will be discussed below.

The outline of this paper is as follows.  In the next section we present
a general discussion of Chern-Simons-like terms in effective actions in
four dimensions, and show that for theories with $N\ge2$ supersymmetry
their existence is a logically pressing issue for carrying out a systematic
derivative expansion.

In sections 3 and 4 we carry out a search for such Chern-Simons-like 
terms in $N=2$ theories, following two different algebraic strategies
which are outlined in section 2.  The results are partial and mainly
negative, except for one class of 4-derivative terms found in section 3
which is superspace Chern-Simons-like only globally on
the Coulomb branch of $N=2$ theories.  

Finally, in section 5 we conclude with some
comments on the new term found in section 3.  It corresponds
to a Dolbeault cohomology class on the Coulomb branch, some
examples of which are given.   For the reader who wishes
to see only the new term, the negative results 
of sections 3 and 4 can probably be skipped without much loss
of comprehensibility.
Section 5 also discusses a class of holomorphic 4-derivative terms
which may exist by 
virtue of other global obstructions on the Coulomb branch.

\section{Derivative expansions, gauge invariance, and extended
supersymmetry}

How we assign weights to the fields is of central importance.
This assignment must be compatible with any global symmetries.
In particular, if space-time derivatives have weight $+1$, then the
supersymmetry algebra implies that the supercharges must be assigned
weight $+1/2$, fixing in turn the relative weights of fields within
the same supermultiplet in supersymmetric theories.
Thus, for $N=1$ supersymmetry, if a scalar is assigned weight $w(\phi)$,
then its fermionic partner in the chiral multiplet must have weight
$w(\phi)+1/2$;
likewise, if the gauge potential has weight $w(A_\mu)$, then the gaugino
will have weight $w(A_\mu)+1/2$.  For $N=2$ supersymmetry, the hypermultiplet
weights are as in the $N=1$ chiral multiplet, while the scalar and vector
fields
in the vector multiplet must have the same weight $w(\phi)=w(A_\mu)$, and
the fermions weight greater by $1/2$. $N=4$ supersymmetric theories
have the same weight assignment as in the $N=2$ vector multiplet.

In theories with a moduli space of inequivalent vacua labelled
by the expectation values of scalar fields, one must assign weight
$0$ to the scalars in order to study the effective action as a function
on the moduli space.  This weight assignment, which is not the same
as the canonical scaling dimensions of the fields, has been used
repeatedly in studies of effective actions with extended supersymmetries
\cite{gsw87,h9507,ds9705,aabe0306}.  It implies, in particular, that
for $N\ge 2$ theories $w(\phi)=w(A_\mu)=0$, and $w(\psi_\alpha)=1/2$,
leading to the problem of characterizing or disproving the existence
of Chern-Simons-like terms, as explained above.

In a superfield formalism, the existence of Chern-Simons-like
terms is more subtle.  Superfields are needed to carry out general
derivative expansions while preserving supersymmetry.  For in
an on-shell and/or component formalism, systematic expansions
become very difficult because one must self-consistently correct
the supersymmetry transformation rules order by order in the
derivative expansion at the same time that one tries to construct
the supersymmetry invariant higher-order terms in the action.  In
an off-shell superfield formulation, though, the supersymmetry
transformations are independent of the form of the action.  In
this case, it only remains to list all the supersymmetry
invariants with a given number of derivatives.  A prescription for
generating all possible such terms might only exist if the
superfields are unconstrained. The unconstrained superfield
formulation of $N=1$ supersymmetry is familiar (see \eg,
\cite{wb92}), while harmonic superspace \cite{gikos84} gives such
an unconstrained formulation for $N=2$, and $3$ supersymmetries
(see \eg, \cite{gios01}).\footnote{For the $N=2$ vector multiplet, 
which will be the focus of this paper, other unconstrained
superfield formalisms exist:  $N=2$ global superspace
\cite{gsw78} with unconstrained real potential superfield \cite{m79,s79}
of derivative weight $w(V_{(ij)})=-3$ related to the field strength 
by $W=\Db^4 D^i\cdot D^j V_{ij}$; or projective superspace
\cite{klr84,lr87} with unconstrained analytic potential superfield
\cite{lr90} of derivative weight $w(V)=-1$ related to the field
strength by $4\Wb=\oint d\zeta \triangle^2 V$.  Both these
formalisms suffer from the same problem of negative derivative
weight potential superfields as the harmonic superspace
formalism.  It is algbraically more complicated to search for 
Chern-Simons-like terms in the global superspace formalism 
because of the potential's more negative weight.  The projective
and harmonic formalisms turn out to be equivalent in
their algebraic complexity \cite{k9806}.}

Superspace Chern-Simons-like terms are gauge-invariant
terms in the action which cannot be written solely in terms of the
field-strength superfield and derivatives, but must also include at
least one vector potential superfield.  For example, in the $N=1$ superspace
description of a $\U(1)$ gauge theory, the vector potential superfield
is the real, gauge-variant, field $V$, while the field strength
superfield is the chiral $W_\a = -{1\over4}\Db^2 D_\a V$.
The question of the existence of superspace Chern-Simons-like terms 
in this theory is then whether there are gauge-invariant terms of the 
form $\int d^4xd^4\th\,f(V,D_\a,\Db_\ad)$ which cannot be rewritten as 
$\int d^4x d^4\th\, g(W_\a,\Wb_\ad,D_\a,\Db_\ad)$ by integration
by parts in $x$ or $\th$ (or similarly for chiral terms integrated
over only half of superspace).

Superspace Chern-Simons-like terms are
a logically broader category than Chern-Simon-like terms:
expansion of a superspace Chern-Simons-like term in component
fields need not give rise to a Chern-Simons-like term for
the component gauge fields.  Indeed, as mentioned above, it
is believed that such Chern-Simons-like terms do not exist in
even space-time dimensions.  On the other hand, superspace Chern-Simons-like
terms are, in fact, known to exist;  the 2-derivative (kinetic) terms
of the $N=3$ supersymmetric $\U(1)$ theory \cite{gikos85} are given as
superspace Chern-Simons terms in $N=3$ harmonic superspace
\cite{r83,gikos85,gio87,gios01}.

In this paper we will initiate a search for such superspace Chern-Simons-like
terms in supersymmetric theories.  By the scaling argument discussed above,
this is a logically pressing issue for making a systematic derivative
expansion on the Coulomb branch of $N=2$ theories.  It is more a
matter of curiosity whether such terms exist in $N=1$ superspace,
so we will only comment briefly on the $N=1$ case in what follows.
In either case, the existence of superspace Chern-Simons-like terms
is a difficult algebraic question.

There are two broad strategies we pursue to search for superspace
Chern-Simons-like terms.  We can use \\
\indent $\bullet$ gauge-variant (vector potential) superfields, or\\
\indent $\bullet$ gauge-invariant (field strength) component fields.\\
%\indent $\bullet$ instanton contributions.\\
Each of these strategies has its limitations which we
now describe.

The gauge-variant superfield strategy, pursued in section 3 below,
is the straight-forward search for gauge invariant terms in the
action involving vector multiplets of the form (schematically)
\be\label{genvecterm} S = \int\! d\zeta\, f(V, D) , \ee which
cannot be rewritten in the form \be S = \int\! d\zeta\, g(W,D) .
\ee Here $d\zeta$ is the appropriate superspace measure, $D$
denotes all the various superspace covariant derivatives, $V$
denotes the potential superfield, and $W$ the field strength
superfield.  For instance, for $N=1$ supersymmetry, $V$ is a real
scalar superfield, and $W_\a$ is a chiral spinor superfield, while
for $N=2$ in harmonic superspace $\Vpp$ is a real analytic scalar
superfield, while $W$ is a chiral scalar superfield. (The
following arguments work for both $N=1$ and $N=2$ supersymmetry,
so we drop the indices on $V$ and $W$.)  The problem with this
strategy is that at a given order in the derivative expansion an
arbitrary number of $V$'s can enter since they have non-positive
derivative weight, $w(V)\le 0$.  To make progress, we then
organize our search by looking for superspace Chern-Simons-like
terms that can be written with only a set number, $\ell$, of
$V$'s, schematically: \be\label{specvecterm} S = \int\! d\zeta\,
V^\ell F(W, D) . \ee In section 3 we carry this out for $\ell=1$.
At $\ell=2$ such a direct search is already algebraically
prohibitively complicated.  We will find, though, an interesting
$\ell=1$ term which is Chern-Simons-like \emph{globally} on the
Coulomb branch of $N=2$ theories.

The second strategy makes the simplifying assumption that there are
no Chern-Simons-like terms (as opposed to \emph{superspace}
Chern-Simons-like terms) in even dimensions, so that in components,
every term can be written in terms of field-strengths $F_{\mu\nu}$
without any explicit gauge potentials $A_\mu$.  Indeed, a partial
fixing of the gauge invariance for either $N=1$ or $N=2$ vector
multiplets allows us to set all but a finite number of auxiliary
fields to zero, leaving the gauge-variant vector potential, $A_\mu$,
as well as gauge invariant scalars and spinors, which we'll
collectively denote by $\phi$, as component fields.  In this gauge
the general term in the action (\ref{genvecterm}) becomes
\be\label{genvecterm2}
S = \int\! d^4x\, g(A_\mu, \phi, \del_\nu),
\ee
where $g$ is Lorentz invariant and gauge invariant under $\de A_\mu =
\del_\mu \ell$.  Since the $\phi$'s are gauge invariant and assuming
that there are no Chern-Simons-like terms in even dimensions, it follows
that up to total derivatives (\ref{genvecterm2}) can be written as
\be\label{genvecterm3}
S = \int\! d^4x\, h(F_{\mu\nu}, \phi, \del_\rho) .
\ee
Thus we can search for superspace Chern-Simons-like terms making
no assumptions on the number of factors, $\ell$, of $V$ that
appear by abandoning
superfields and working in terms of gauge-invariant component
fields.  We carry this out in section 4 to show that there
are no 3-derivative superspace Chern-Simons-like terms on
a one-dimensional $N=2$ Coulomb branch.  The price we pay is
that since we are working in components, we have to check
supersymmetry ``by hand".

It may be helpful at this point to remark on a connection between such
superspace Chern-Simons-like terms and the issue of locality
in the Grassmann coordinates of superspace.
Since $F_{\mu\nu}$, $\phi$, and their derivatives are
just components of the field strength superfield $W$ and
its derivatives, we can write (\ref{genvecterm3}) as
\be\label{genvecterm4}
S = \int\! d^4x\, h\left(\textstyle{\int\!d\th_1 j_1(W,D)\,,
\,\int\!d\th_2 j_2(W,D)\,,\ldots}\right) ,
\ee
where the $j_n$ are arbitrary functions of superspace covariant
derivatives and $W$'s, and the $d\th_i$ are appropriate integration
measures over the Grassmann-odd superspace coordinates.  Thus we have
rewritten the general vector multiplet term (\ref{genvecterm}) solely in
terms of the field strength superfield.
But (\ref{genvecterm4}) is not local in superspace.  Such a superspace-local
term would have just a single integral over the Grassmann-odd coordinates,
$S_{\mbox{local}} = \int\! d^4x\,d\th \, h(W,D)$.  Fayet-Iliopoulos terms
provide a simple example.  Fayet-Iliopoulos terms are
superspace Chern-Simons-like terms, in the sense that they cannot
be written in terms of field strength superfields integrated over
the usual superspace.  For example, the $N=1$ Fayet-Iliopoulos terms
$\int d^4xd^4\th\,V$ is gauge-invariant and cannot be written in terms of
$W_\a$ integrated over the full superspace $\int d^4xd^4\th$ or chiral
superspace $\int d^4x_Cd^2\th$.  However, by the above argument it
can be written in terms of $W_\a$ integrated over some other superspace.
Indeed, it is given by an integral over one quarter of superspace, $\int
d^4xd\th^\a\, W_\a$, and is supersymmetric by virtue of the Bianchi
identity that $W_\a$ satisfies.  Unlike (\ref{genvecterm4}) it is local
in superspace, but only because it involved only a single field.
See \cite{aabe0306} for a more detailed discussion of superspace
locality.

A third strategy for finding superspace Chern-Simons-like terms,
which we hope to report on elsewhere \cite{ab0410},
is to use an instanton calculation to show the existence of
certain component terms in the action which are known not to arise
from any supersymmetric term involving just field strength
superfields. 
%In particular, it can be shown that a derivative
%weight +4 term involving the product of eight left-handed fermions
%cannot arise in the $\U(1)$ effective actions of $N=1$ or $N=2$
%theories using only field strength superfields, \eg, using the
%classification of dimension-4 terms found in \cite{aabe0306} for
%the $N=2$ case.  A one instanton contribution on the Coulomb branch
%of an $N=4$ theory naively gives rise to such an 8-fermion just by
%the usual counting of fermionic zero-modes, though calculation of
%its coefficient is tricky due to potential cancellations \cite{ds9705}.  
The main drawback of this strategy is that it is
not systematic, so cannot rule out the existence of general
Chern-Simons-like terms, but only of certain very special ones.

\section{Gauge-variant superfield arguments}

In the rest of this paper we work on the Coulomb branch of an $N=2$ gauge
theory where the low energy effective action at a generic vacuum includes
only massless $\U(1)$ vector multiplets and massless neutral hypermultiplets.
Furthermore, for simplicity we will ignore the hypermultiplets, and only
consider terms with vector superfields.
The propagating component fields of a $\U(1)$ vector superfield
are massless neutral
scalars and spinors, $\phi$, $\psi_\a$, and $\U(1)$ vectors, $A^\mu$.
In harmonic superspace the vectormultiplet is represented either
by the potential superfield $\Vpp$ or the field strength superfield
$W$.

\paragraph{Harmonic superspace:}
We now very briefly review the salient points of the harmonic
superspace formalism concerning vector superfields.  We follow
the notation of \cite{gios01} where a detailed exposition
of harmonic superspace can be found; a concise review appears
in \cite{aabe0306}.

An important feature of harmonic superspace is that, in addition to the
usual space-time directions described by coordinates $x^\mu$ and
Grassmann-odd directions with spinor coordinates, $\th^\pm_\a$ and
$\thb^\pm_\ad$,
there is also a 2-sphere described by commuting harmonic coordinates
$u^\pm_i$, $i\in\{1,2\}$.  Expansion in the $u$'s gives rise to an infinite number
of auxiliary fields.  Though terms in the effective action need
not be local in the $u$'s, there exists a systematic procedure
to list all such terms \cite{aabe0306}.  In the case of
vector superfields, we will see that the $u$'s play only a minor role.

Superspace covariant derivatives, $D^\pm_\a$ and $\Db^\pm_\ad$,
are introduced in the usual way, along with a set of covariant
$u$ derivatives denoted
$\Dpp$, $\Dmm$, and $D^0$.  The $u$-derivatives satisfy an
$\SU(2)_R$ algebra
\be\label{su2alg}
[D^0,D^{\pm\pm}] = \pm 2D^{\pm\pm}, \qquad [\Dpp,\Dmm] = D^0,
\ee
while the covariant derivatives
satisfy the $N=2$ algebra
\be\label{Dalg}
\{D^\pm_\a,\Db^\mp_\ad\} = \mp 2i\dsl_{\a\ad} ,\qquad
{}[D^{\pm\pm},D^\mp_\a] = D^\pm_\a, \qquad
{}[D^{\pm\pm},\Db^\mp_\ad] = \Db^\pm_\ad,
\ee
with all other (anti)commutators vanishing.  Eqs.\ (\ref{Dalg}) and
(\ref{su2alg}) give the form of the $N=2$ algebra on harmonic
superspace that we will use.
$N=2$ supersymmetry invariants can be formed by integrating a general
harmonic superfield over all the superspace coordinates with measure
$\int du\, d^4x\, d^4\thp\, d^4\thm$, where $du$ is the appropriate
measure for integration over the $u$-sphere.

The $\pm$ superscripts denote the charge under $\U(1)_R\subset
\SU(2)_R$.  $N=2$ invariant terms are required to be neutral
under this $\U(1)$.  Also, all functions of the $u^\pm$ are
required to be harmonic, which is to say that they have
regular power series expansions in the $u^\pm$.

Two different constraints, the {\em chiral constraint} and the
{\em analytic constraint}, can
be used to reduce superfield representations in $N=2$ harmonic superspace.
The chiral constraint,
\be\label{CC}
\Dbp_\ad W = \Dbm_\ad W = 0,
\ee
is solved by introducing a chiral space-time coordinate $x^C$
annihilated by $\Db^\pm$.  Then the chiral constraint
is solved by an arbitrary (unconstrained) superfield independent
of the $\thb^\pm$'s: $W = W(x_C^\mu,\th^\pm_\a,u^\pm_i)$.  The
field-strength superfield for the vector multiplet is such a
chiral superfields.  Supersymmetry invariants
can be constructed by integrating chiral superfields against the measure
$\int\! du\, d^4x_C\, d^4\th = \int\! du\, d^4x\, D^4$,
where $D^4 \equiv {1\over16} (\Dp)^2 (\Dm)^2$.
The analytic constraint,
\be\label{AC}
\Dp_\a V = \Dbp_\ad V = 0,
\ee
is solved by introducing an analytic space-time coordinate $x_A$
annihilated by $\Dp$ and $\Dbp$, so that an arbitrary (unconstrained)
superfield independent of $\thm$ and
$\thbm$, $V = V(x_A^\mu,\thp_\a,\thbp_\ad,u^\pm_i)$, solves the
analytic constraint.  These {\em
analytic superfields} are useful for describing the vector potential
superfield.
Supersymmetry invariants can be constructed by integrating analytic
superfields against the measure
$\int\! du\, d^4x_A\, d^4\thp = \int\! du\, d^4x\, (\Dm)^2 (\Dbm)^2$.
Note, in particular, that $d^4\thp$ has $\U(1)_R$ charge $-4$ because
Grassmann integration is differentiation.

The unconstrained $N=2$ vector multiplet superfield is a
(real) analytic (\ref{AC}) superfield $\Vpp$ transforming
under $\U(1)$ gauge transformations as
\be\label{devpp}
\de\Vpp = - \Dpp \lambda,
\ee
where $\lambda$ is an arbitrary real analytic superfield.
The gauge invariant field strength superfield is constructed as follows.
First, another gauge potential superfield $\Vmm$ is defined in terms
of $\Vpp$ as the solution to the differential equation in $u^\pm$
\be\label{Vmmdef}
\Dpp \Vmm=\Dmm \Vpp ,
\ee
which has a unique solution by virtue of the harmonicity requirement on
the $u$-sphere.  $\Vmm$ is not an analytic (or anti-analytic)
superfield, but is real and transforms under gauge
transformations as $\de\Vmm = -\Dmm\lambda$.
Two useful identities involving $\Vmm$ are
\be\label{useful}
D^-_\a\Vpp = - \Dpp D^+_\a \Vmm,\qquad
D^-_\a\Vmm = - \Dmm D^+_\a \Vmm,
\ee
and similarly with $\Db$'s.
The field strength superfield is then defined by
\be\label{Wdef}
W=-{1\over 4 }(\Dbp)^2 \Vmm .
\ee
It is a straight forward exercise, using the $N=2$ algebra (\ref{su2alg})
and
(\ref{Dalg}), to check that $W$ is gauge invariant, chiral (\ref{CC}),
$u$-independent
\be
D^{\pm\pm} W = 0 ,
\ee
and satisfies the Bianchi identities
\be\label{bianchi}
D^\pm\cdot D^\pm W = \Db^\pm\cdot\Db^\pm \Wb ,\qquad
D^\pm\cdot D^\mp W = \Db^\pm\cdot\Db^\mp \Wb .
\ee
The $u$-independence of $W$ implies that
in expressions involving the field strength superfields alone
(\ie, no $V^{\pm\pm}$'s), the integration over the auxiliary $u$-sphere
can be done separately, leaving an expression in standard $N=2$ superspace
with coordinates $\{x^\mu, \th^\pm_\a, \thb^\pm_\ad\}$.

The lowest component of $W$ is the
complex scalar $\phi$ whose vevs parameterize the Coulomb branch.
Thus $W$ must be assigned derivative weight $w(W)=0$.  Since
$w(D)=1/2$ and the $u$'s and therefore the $D^{\pm\pm}$ derivatives
have weight $0$, (\ref{Wdef}) and (\ref{Vmmdef}) imply that
$w(V^{\pm\pm})=-1$.  This negative weight is the source of the
problem of Chern-Simons-like terms in $N=2$ effective actions.

\paragraph{Terms with no $V$s:}
Before starting our search for harmonic superspace
Chern-Simons-like terms, we first review the classification of
Coulomb branch terms that can be written solely using the field
strength superfield and its derivatives.  The complete set of such
terms up to four derivatives is \cite{aabe0306} \bea\label{Wterms}
S_1 &=& \int\! d^4x\, d\th^i\cdot d\th^j\, \xi_{ij}\, W ,
\qquad i,j\in\{1,2\},\quad \xi_{ij}\in\R,\nonumber\\
S_2 &=& \int\! d^4x\, d^4\th\ {\cal F}(W) +\mbox{c.c.} ,
\nonumber\\
S_{4a} &=& \int\! d^4x\, d^4\th \,\,
\del_\mu W \del^\mu W \, {\cal G}(W) +\mbox{c.c.},
\nonumber\\
S_{4b} &=& \int\! d^4x\, d^4\th\, d^4\thb \,\,
{\cal H}(W,\Wb) .
\eea
The 1-derivative term is the Fayet-Iliopoulos term;
though an integral over only 1/4 of superspace, it
is $N=2$ invariant by virtue of the extra constraint (\ref{bianchi})
that $W$ satisfies.  The 2-derivative term is
the well-known holomorphic prepotential term, encoding
generalized kinetic, Yukawa, and $\psi^4$
terms.  There are no 3-derivative terms, and two
independent 4-derivative terms, the first of which is
holomorphic; note that when there is only a single vector
multiplet, $S_{4a}$ can be rewritten using the Bianchi identity
as an $S_{4b}$ term \cite{aabe0306}.  These terms will be
discussed in more detail with a view towards possible
global obstructions to their existence in section 5 below.

Now we turn to the question of whether there exist gauge
invariant $\U(1)$ vector multiplet terms which cannot be written solely
in terms of the field strength multiplets $W$.  Let us examine
this possibility by assuming that such a term can be written
with just one power of the potential superfield, \ie, schematically
of the form
\be\label{oneV}
S = \int\!d\zeta\, V^{\pm\pm}\, f(W,D),
\ee
where $d\zeta$ is some
superspace measure and $f$ is an arbitrary function of field strength
superfields and covariant derivatives.  (Note that, by integration
by parts, we can always write such terms with no derivatives acting
on $V$.)  We must first determine the conditions on $f$ such
that $S$ is gauge invariant.  Then we must show that it cannot
be written as one of the terms in (\ref{Wterms}).  Only then
will we have found a superspace Chern-Simons-like term.

There are two possible choices for the measure $d\zeta$ in (\ref{oneV}).
Since $\Vpp$ is analytic, $S$ could be $N=2$ invariant if $d\zeta=dud^4x
d^4\thp$, the integration over the analytic half of superspace.  The
other possibility is that $d\zeta$ could be the integration measure over
all of superspace.  We will explore these two possibilities in turn.

\paragraph{One-derivative terms with one $V$:}
If the integration is only over analytic superspace, the integrand
in (\ref{oneV}) must be analytic superfield.  This limits its form
to \be S_A = \int\!d^4x du d^4\thp\, f^{++}_a \Vpp_a \ee where $a$
is an index labelling different $\U(1)$ vector multiplets, and
$f^{++} = f^{++}(u^\pm, D^{\pm\pm},\del_\mu)$ is an arbitrary
function of $u$ and $u$- and $x$-derivatives; in particular, no
dependence on $W$ or $\Wb$ is allowed by analyticity.  Since the
derivatives can be taken to not act on $\Vpp$ by integration by
parts, we can drop the derivatives altogether, and
$f^{++}=f^{++}(u^\pm)$.  Thus $S_A$ can only be a 1-derivative
term since $w(\Vpp)=-1$ and $w(d^4\thp)=+2$. The gauge variation
of $S_A$ is then \be \de S_A = -\int\!d^4x du d^4\thp\,
f^{++}(u)\cdot \Dpp \la = \int\!d^4x du d^4\thp\, \left[\Dpp
f^{++}(u)\right] \cdot \la \ee where in the second step we have
integrated by parts, and where we have suppressed the $a$ index,
using $\cdot$ to denote contraction over these flavor indices.
Since $\la$ are arbitrary analytic superfields, gauge invariance
implies that $\Dpp f^{++} = 0$, or that $f^{++}$ is independent of
$u^-$.  In order to have total $\U(1)_R$ charge $+2$, we therefore
have $f^{++}(u) = u^+_i u^+_j \xi^{ij}$, where $\xi^{ij}$ are some
constants.  Now, $d^4\thp = d^2\thp (\Dbm)^2$, so \be\label{sa5}
S_A = \int\!d^4x du d^2\thp\, u^+_i u^+_j \xi^{ij} (\Dbm)^2\Vpp.
\ee {}From (\ref{useful}), (\ref{Dalg}), and (\ref{Wdef}) it
follows that $(\Dbm)^2\Vpp = -\Dpp\Dbm\cdot\Dbp\Vmm -4W$.
Inserting this into (\ref{sa5}), the first term vanishes after
integration by parts, leaving $S_A \propto \int\!d^4x d\th_i\cdot
d\th_j \si^{ij} W$, which is just the Fayet-Iliopoulos term $S_1$
in (\ref{Wterms}).

We now turn to the terms written as integrals over the full superspace.
Since $w(d^8\th)=4$ and $w(V)=-1$, these terms have at least 3 derivatives.
We will examine only the 3- and 4-derivative terms found this way.

\paragraph{Three-derivative terms with one $V$:}
The general 3-derivative term is \be\label{s3a} S_3 =
\int\!d^4xdud^8\th\,\left\{ g^{(-2)}(W,\Wb,u) \cdot \Vpp +
g^{(+2)}(W,\Wb,u)\cdot\Vmm \right\}. \ee Now, any function
$g^{(+2)}$ can be written as $g^{(+2)}=\Dpp g^{(0)}$ for some
$g^{(0)}$ by harmonicity in $u$.  Then $\int\!du\, g^{(+2)}\Vmm =
\int\!du\, (\Dpp g^{(0)})\Vmm = -\int\!du\, g^{(0)}\Dpp\Vmm =
-\int\!du\, g^{(0)}\Dmm\Vpp = \int\!du\, (\Dmm g^{(0)})\Vpp =
\int\!du\, \tilde g^{(-2)}\Vpp$. Thus the $\Vmm$ term can be
converted to the $\Vpp$ term, and so can be dropped from
(\ref{s3a}).  The gauge variation of $S_3$ after integration by
parts is \be\label{s3b} \de S_3 = \int\!d^4xdud^8\th\,\left[ \Dpp
g^{(-2)} \right] \cdot \la , \ee which vanishes if and only if
\be\label{s3c} \Dpp g^{(-2)} = \Dp \bar h^{(-1)} + \Dbp h^{(-1)}
\ee for some $\bar h^{(-1)}$ and $h^{(-1)}$, since $\la$ is
analytic (\ie, annihilated by $\Dp$ and $\Dbp$). Since the left
hand side of (\ref{s3c}) is a function of $W$ and $\Wb$ only, we
must have $\bar h^{(-1)} = \bar f(\Wb,u)\Dp\Vmm$ and $h^{(-1)} =
f(W,u)\Dbp\Vmm$ for some $f$ and $\bar f$. That then implies that
$\Dpp g^{(-2)}$, and therefore $g^{(-2)}$, is a sum of a function
of $W$ alone and of $\Wb$ alone.  Thus, for gauge invariance, we
have \be\label{s3d} S_3 = \int\!d^4xdud^8\th\, \Vpp\cdot
\left\{g^{'(-2)}(W,u)+\bar g^{'(-2)}(\Wb,u)\right\}. \ee But
$\int\!d^8\th = \int\!d^4\thp(\Dp)^2 (\Dbp)^2$, and since both
$\Dp$ and $\Dbp$ annihilate $\Vpp$ (by analyticity), and one or
the other of them annihilate $g'$ or $\bar g'$ (by chirality of
$W$ and $\Wb$), $S_3$ vanishes.  Thus there are no 3-derivative
terms with a single $V$.

\paragraph{Four-derivative terms with one $V$:}
The most general such expression has many terms:
\bea\label{s4a}
S_4 &=& \int\!d^4xdud^8\th\,\Biggl\{
\Vpp_a\Bigl[ g^{(0)bc}_a \Dm W_b\cdot \Dm W_c
+ \til g^{(-2)bc}_a \Dm W_b\cdot \Dp W_c
+ \til g^{(-4)bc}_a \Dp W_b\cdot \Dp W_c \nonumber\\
&& \qquad\qquad\qquad\qquad\mbox{}+
f^{(0)b}_a (\Dm)^2 W_b
+ f^{(-2)b}_a (\Dm\Dp) W_b
+ f^{(-4)b}_a (\Dp)^2 W_b \Bigr]  \nonumber\\
&&\qquad\qquad\ \ \mbox{}+
\Vmm_a\Bigl[ h^{(0)bc}_a \Dp W_b\cdot \Dp W_c
+ h^{(2)bc}_a \Dm W_b\cdot \Dp W_c
+ h^{(4)bc}_a \Dm W_b\cdot \Dm W_c \nonumber\\
&& \qquad\qquad\qquad\qquad\mbox{}+
d^{(0)b}_a (\Dp)^2 W_b
+ d^{(2)b}_a (\Dm\Dp) W_b
+ d^{(4)b}_a (\Dm)^2 W_b \Bigr]  \Biggr\} + \mbox{c.c.}
\eea
where the $d$, $f$, $g$, and $h$'s are all functions of
$W$, $\Wb$, and the $u$'s.  Even before requiring gauge
invariance, this can be drastically simplified.  First, consider the
$\Vmm$ terms.  We can write $h^{(4)}=\Dpp\til h^{(2)}$,
$h^{(2)}=\Dpp\til h^{(0)}$, and $h^{(0)}=h+\Dpp\til h^{(-2)}$
for some $\til h$'s, and similarly for the $d$'s, where $h$ is
the $u$-independent piece of $h^{(0)}$.  Then integrate
by parts on $\Dpp$, rewrite $\Dpp\Vmm \to \Dmm\Vpp$, and
integrate by parts on $\Dmm$, to convert the $\til h$ and $\til d$
terms to terms of the same form as the $\Vpp$ terms.  Therefore
these terms can all be dropped.  Next consider the $f^{(-2)}$ and
$f^{(-4)}$ terms.  By redefining $\til g \to g$ by
adding appropriate derivatives of the $f$'s with respect to $W$,
these $f$ terms can be written as total $\Dp$ derivatives times
$\Vpp$.  Integrating by parts on $\Dp$ gives 0 since $\Dp\Vpp=0$
by analyticity.  Therefore we can drop these terms as well.
Thus $S_4$ has been simplified to
\bea\label{s4b}
S_4 &=& \int\!d^4xdud^8\th\,\Biggl\{
\Vpp \Bigl[ g^{(0)} \Dm W\vee \Dm W
+ g^{(-2)} \Dm W\vee \Dp W
+ g^{(-4)} \Dp W\vee \Dp W \nonumber\\
&& \qquad\qquad\qquad\qquad\mbox{}+
e^{(-2)} \Dm W\wedge \Dp W
+ f^{(0)} (\Dm)^2 W \Bigr]  \nonumber\\
&&\qquad\qquad\ \ \mbox{}+
\Vmm\Bigl[ h\, \Dp W\vee \Dp W
+ d\, (\Dp)^2 W \Bigr]  \Biggr\} + \mbox{c.c.}
\eea
where $e^{(-2)}$, $f^{(0)}$, and the $g^{(n)}$ are
functions of $W$, $\Wb$, and the $u$'s; $d$ and $h$
are functions of $W$ and $\Wb$ only; we have
suppressed the flavor indices, which should all
be contracted with indices on the coefficient functions;
we have introduced the
notations $A\wedge B = {1\over2}(A_a B_b-A_bB_a)$
and $A\vee B = {1\over2}(A_a B_b+A_bB_a)$
for antisymmetrized and symmetrized indices respectively;
and $e^{(-2)}$ and $g^{(-2)}$ are introduced as the
antisymmetric and symmetric parts of $\til g^{(-2)}$.

We now demand that $S_4$ be gauge invariant.  Taking
the gauge variation, integrating by parts on $D^{\pm\pm}$ and
on $\Dp$ for the resulting $(\Dm\Dp)W$ term, and collecting terms
gives, after some algebra,
\bea\label{s4c}
\de S_4 &=& \int\!d^4xdud^8\th\,\la \Biggl\{
\left[ \Dpp g^{(0)}\right] \Dm W\vee \Dm W
+ \left[ \Dpp f^{(0)}\right]  (\Dm)^2 W \nonumber\\
&&\qquad\qquad\qquad\mbox{}+
\left[ \Dpp e^{(-2)} - 2\del\wedge(d+f^{(0)})\right]
\Dm W\wedge \Dp W \nonumber\\
&&\qquad\qquad\qquad\mbox{}+
\left[ \Dpp g^{(-2)} - 2\del\vee(d+f^{(0)})+2(h+g^{(0)})\right]
\Dm W\vee \Dp W \nonumber\\
&&\qquad\qquad\qquad\mbox{}
+ \left[ \Dpp g^{(-4)} + g^{(-2)}\right]
\Dp W\vee \Dp W  \Biggr\} + \mbox{c.c.}
\eea
where $\del = \del/\del W$.  Gauge invariance
then implies that the terms in square brackets
must vanish, giving
\bea\label{s4d}
\Dpp g^{(0)} &=& \Dpp f^{(0)}=0,
\nonumber\\
\Dpp e^{(-2)} &=& 2\del\wedge(d+f^{(0)}),
\nonumber\\
\Dpp g^{(-2)} &=& 2\del\vee(d+f^{(0)})+2(h+g^{(0)}),
\nonumber\\
\Dpp g^{(-4)} &=& - g^{(-2)}.
\eea
The first line of (\ref{s4d}) implies that $g^{(0)}=g$ and
$f^{(0)}=f$ are independent of $u$.  Thus the right hand
sides of the second and third lines are $u$-independent,
which then implies that $e^{(-2)}=g^{(-2)}=0$, which in turn
implies $g^{(-4)}=0$ by the fourth line.  Furthermore, the
right hand sides of the second and third lines then vanish,
giving $\del\wedge(d+f)=0$, and $\del\vee(d+f)= g+h$.
Define $\hat g=g-\del\vee f$ and $\hat h = h-\del\vee d$
so that
\bea\label{s4e}
S_4 &=& \int\!d^4xdud^8\th\,\biggl\{
\Vpp \left[ \hat g\, \Dm W\vee \Dm W
+ \Dm(f\, \Dm W) \right]  \nonumber\\
&&\qquad\qquad\ \ \mbox{}+
\Vmm\left[ \hat h\, \Dp W\vee \Dp W
+ \Dp (d\, \Dp W) \right]  \biggr\} + \mbox{c.c.}
\eea
where $d$, $f$, $\hat g$, and $\hat h$ are $u$-independent
functions of $W$ and $\Wb$ satisfying
\be\label{s4f}
\del\wedge(d+f)=0, \qquad\mbox{and}\qquad
\hat g+ \hat h =0.
\ee
This can be simplified further.  Consider the following
manipulation of the $\hat g$ term:
\bea\label{s4g}
\int\!d^4xdud^8\th\,\Vpp\,\hat g\, \Dm W\vee \Dm W
&=& \int\!d^4xdud^8\th\,\Vpp\,\hat g\, (\Dmm\Dp W)\vee \Dm W
\nonumber\\
&=& \int\!d^4xdud^8\th\,\Vpp\,\hat g\, \Dmm(\Dp W\vee \Dm W)
\nonumber\\
&=& -\int\!d^4xdud^8\th\,(\Dmm\Vpp)\,\hat g\, \Dp W\vee \Dm W
\nonumber\\
&=& -\int\!d^4xdud^8\th\,(\Dpp\Vmm)\,\hat g\, \Dp W\vee \Dm W
\nonumber\\
&=& \int\!d^4xdud^8\th\,\Vmm\,\hat g\, \Dpp(\Dp W\vee \Dm W)
\nonumber\\
&=& \int\!d^4xdud^8\th\,\Vmm\,\hat g\, \Dp W\vee (\Dpp\Dm W)
\nonumber\\
&=& \int\!d^4xdud^8\th\,\Vmm\,\hat g\, \Dp W\vee \Dp W,
\eea
where we have used extensively that $W$ is $u$-independent.
This shows that the $\hat g$ term is the same as the $\hat h$
term.  But since $\hat g=-\hat h$ by (\ref{s4f}), they cancel.
Now consider the following manipulation of the $f$ term:
\bea\label{s4h}
\int\!d^4xdud^8\th\,\Vpp\,\Dm(f\, \Dm W)
&=& -\int\!d^4xdud^8\th\,(\Dm\Vpp)\, f\, \Dm W
\nonumber\\
&=& \int\!d^4xdud^8\th\,(\Dpp\Dp\Vmm)\, f\, \Dm W
\nonumber\\
&=& -\int\!d^4xdud^8\th\,(\Dp\Vmm)\, f\, \Dpp\Dm W
\nonumber\\
&=& -\int\!d^4xdud^8\th\,(\Dp\Vmm)\, f\, \Dp W
\nonumber\\
&=& \int\!d^4xdud^8\th\,\Vmm\, \Dp(f\, \Dp W),
\eea
which is of the same form as the $d$ term.  Calling
$A(W,\Wb)\equiv d+f$, and restoring the flavor indices,
the final form for gauge-invariant 4-derivative terms with
one $V$ is
\be\label{s4i}
S_4 = \int\!d^4xdud^8\th\,
V_a^{--}\, \Dp \left(A^b_a(W,\Wb) \Dp W_b\right) + \mbox{c.c.},
\ee
where, from (\ref{s4f}), $A$ satisfies
\be\label{s4j}
\del^c A^b_a - \del^b A^c_a = 0 .
\ee

The next step is to determine when this term can be rewritten
solely in terms of $W$'s and $\Wb$'s.  To do this we need to have
the $\Dp$'s act on $\Vmm$.  But (\ref{s4j}) is precisely the local
integrability condition for \be\label{s4k} A^b_a = \del^b B_a \ee
for some $B_a(W,\Wb)$.  In this case $A^b_a \Dp W_b = \del^b B_a
\Dp W_b = \Dp B_a$, and $S_4$ becomes \be\label{s4l} S_4 =
\int\!d^4xdud^8\th\, V_a^{--}\, (\Dp)^2 B_a = \int\!d^4xdud^8\th\,
B_a\, (\Dp)^2 V_a^{--} = -4\int\!d^4xd^8\th\, B_a\, \Wb^a, \ee
written solely in terms of field strength superfields.  However,
this rewriting was possible only locally on the Coulomb branch,
since globally there might be an obstruction to the integrability
of (\ref{s4k}).  Thus (\ref{s4i}) may, in fact, be a superspace
Chern-Simons-like term, albeit only globally on the moduli space.
We will return in section 5 to discuss the existence and implications of
such terms.

Aside from discovering this new term, there are few general
lessons we can extract from this calculation.  Since
it assumed a specific form, namely only one explicit
power of $V^{\pm\pm}$, it allows no general statements
to be made about the existence of Chern-Simons-like terms
at a given derivative order.  Because even with one power
of $V$ the calculation was so algebraically complex, it
seems unlikely that this strategy can be usefully
extended to a general argument valid for all powers
of $V$.  Indeed, even at $V^2$ the algebra is
prohibitively complicated.  As a small example, consider
the following 3-derivative term with two $V$'s:
\be\label{s3V2}
S_3 = \int\!d^4xdud^8\th\,\left[f_{ab}(W,\Wb)+g_{ab}(\Wb,u)\right]\,
\Dp V^{--}_a \Dm V^{++}_b,
\ee
where both $f$ and $g$ are symmetric on $a$ and $b$.
We leave as an exercise for the masochistic reader to show, first,
that this term is gauge invariant, and second, that it actually vanishes.

{}For these reasons, we now turn to a more systematic
approach to the search for superspace Chern-Simons-like
terms.

\section{Gauge-invariant component arguments}

We now search for superspace Chern-Simons-like terms by
looking directly at the component expansion of the vector
multiplet.  As discussed in section 2, assuming that there
are no Chern-Simons-like terms (as opposed to
\emph{superspace} Chern-Simons-like terms) in even dimensions,
any gauge invariant term can be written in terms of the
components of the field strength vector multiplet.

The vector multiplet contains scalar fields  $\phi$ and $\bar\phi$, with derivative
weight zero;  a spinor field $\psi^\alpha$, with derivative weight $1/2$; a triplet
of auxiliary scalar fields $D^{++},D^{--}$, and $D^{+-}$, with derivative
weight one; and $\U(1)$ gauge field strengths
$F^{(\a\b)}$ and $\bar F^{(\ad\bd)}$, symmetric on the spinor indices\footnote{Or, equivalently, anti-symmetric on spacetime indices.} also with
derivative weight one.
In addition to these fields, we may have spacetime derivatives, carrying
derivative weight one.
We write these derivatives contracted with a pauli matrix, giving them a dotted and
an undotted index: $\dsl_{\a\ad} =\sigma^\mu_{\a\ad}\; \del_\mu$.
These components are related to the field strength chiral superfield $W$
and its conjugate $\Wb$ by
\bea\label{Wcomp}
\phi&=&W|_{\theta=\bar\theta=0}\nonumber\\
\bar\phi&=&\bar W|_{\theta=\bar\theta=0}\nonumber\\
\psi^\pm_\a&=&D^\pm_\a W|_{\theta=\bar\theta=0}\nonumber\\
\bar\psi^\pm_\a&=&\bar D^\pm_\ad \bar W|_{\theta=\bar\theta=0}\nonumber\\
D^{\pm\pm}&=&D^{\pm\a}D^\pm_\a W|_{\theta=\bar\theta=0}\;\;
= \bar D^{\pm\pm}
\nonumber\\
D^{+-}&=&D^{+\a}D^-_\a W|_{\theta=\bar\theta=0}\;\; = \bar D^{+-}\nonumber\\
F_{(\a\b)}&=&(D^{+\a}D^{-\b} + D^{+\b}D^{-\a})
W|_{\theta=\bar\theta=0}\nonumber\\
\bar F_{(\ad\bd)}&=&(\bar D^{+\ad}\bar D^{-\bd} + \bar D^{+\bd}\bar D^{-\ad}) \bar W|_{\theta=\bar\theta=0}  .
\eea
Any expression written in terms of these fields will be automatically
gauge invariant, but not manifestly $N=2$ supersymmetric.

We can now organize the search for superspace Chern-Simons-like
term order-by-order in the derivative expansion.  The
non-Chern-Simons-like terms with four or fewer derivatives were
listed in section 3, (\ref{Wterms}). It is easy to see in
components \cite{gsw78,st83} that all 1- and 2-derivative terms
are included in this list.  Since there are no 3-derivative terms
in this list, if we find any in the component method, we will have
then found a superspace Chern-Simons-like term.

We will now show that there are no gauge invariant $N=2$ supersymmetric
3-derivative terms with just a single vector multiplet.  We look at only one 
$\U(1)$ vector multiplet for simplicity.  We will comment on the extension to 
many $\U(1)$'s below.

Our strategy for showing that there are no 3-derivative terms is
to look at a possible term in the action, find its supersymmetry
variation, and then show that this variation cannot be cancelled.
Once this term is shown not to contribute, we then look at another
action term, and so on, until all are exhausted.  The key to doing
this efficiently is to eliminate the terms in a particular order.
We have found that it is convenient to organize the terms by
decreasing number of $\U(1)$ field strength fields $F$. This is
both because the $F$ fields can only be obtained in one way in a
supersymmetry variation, and also because of the limited number of
ways in which they can be included, due to their two spinor
indices and Lorentz invariance.

Note that an action will be supersymmetric not only if the
variation of the Lagrangian vanishes, but also if it is just a
total derivative.  So, it is possible that the variations of
combinations of terms do not cancel, but add to form a total
derivative.   In order for this to happen, the terms must all have
the same fields and Lorentz structure, but with derivatives acting
on different fields.  Since we will almost never have to keep
track of where derivatives act\footnote{Though we do need to keep
track of where derivatives act when looking at terms with one $F$
and no fermions, but only for the purpose of finding out what
fields we need in such a term.  The possibility of a total
derivative does not enter there.}, when we say that  terms cannot
cancel, then they also cannot add to form a total derivative.

$N=2$ supersymmetry is preserved when the action is invariant under four
independent supersymmetry transformations generated by
\be
D^+_\a,\ D^-_\a,\ \bar D^+_\ad,\ \bar D^-_\ad.
\ee
For our purposes checking the supersymmetry variation under one
of these will be equivalent to checking the variation with respect to the
others.  For definiteness we usually look
at the $D^+_\a$ variation, and unless otherwise specified this will be what
we mean by \emph{the} supersymmetry variation.

The supersymmetry transformations of the component fields are
\bea
D^\pm_\a \phi &=& \psi^\pm_\a \nonumber\\
D^\pm_\a \bar\phi &=& 0 \\
\nonumber\\
D^\pm_\a \psi^\pm_\b &=& 1/2 \,\epsilon_{\a\b} D^{\pm\pm}\nonumber\\
D^\pm_\a \psi^\mp_\b &=& \pm \,F_{(\a\b)} + 1/2\, \epsilon_{\a\b}D^{+-}\nonumber\\
D^\pm_\a \bar\psi^\pm_\bd &=& 0\nonumber\\
D^\pm_\a \bar\psi^\mp_\bd &=& \mp\, 2i\, \dsl_{\a\bd}\bar\phi \\
\nonumber\\
D^\pm_\a D^{\pm\pm} &=& 0 \nonumber\\
D^\pm_\a D^{\mp\mp} &=& \mp \,4i\, (\dsl \bar\psi^\mp)_\a \nonumber\\
D^\pm_\a D^{+-} &=& \mp \,2i \,(\dsl \bar\psi^\pm)_\a \\
\nonumber\\
D^\pm_\a F_{(\b\ga)} &=& 4i \,\epsilon_{\a(\b}(\dsl \bar\psi^\pm)_{\ga)} \nonumber\\
D^\pm_\a \bar F_{(\bd\gd)} &=& \mp \, 2i \, \dsl_{\a (\bd}\bar\psi^\pm_{\gd)}
\eea
as is easily read off from the superfield expressions (\ref{Wcomp}) and
the $N=2$ algebra (\ref{Dalg}).

\paragraph{Terms with three $F$'s:}
First of all, we note that with only one (or two) distinct vector multiplets, we cannot
have a term in the action with three $F$ fields, by Lorentz invariance and
symmetry on spinor indices.  Since each $F$ has derivative weight one, there
can be no spinors $\psi$ in a 3-$F$, 3-derivative term, and so the three $F$s'
spinor indices must be contracted.  Because the $F$'s are symmetric on their
spinor indices (and because spinor index contraction is antisymmetric) this trace
is necessarily equal to its negative and therefore zero.  Similarly there can be
no action terms with three $\bar F$'s, and we cannot make a derivative weight
three Lorentz scalar that contains both $F$ and $\bar F$.

\paragraph{Terms with two $F$'s:}
Now we look at action terms with two $F$ fields.  The only
possible 3-derivative Lorentz scalars are \be\label{2f}
(\psi^\pm)^2 \;tr(F^2),\qquad (\bar\psi^\pm)^2 \;tr(F^2),\qquad
D^{\pm\pm}\; tr(F^2), \ee as well as their conjugates, involving
$\bar F^2$. These terms can have arbitrary coefficients which are
functions of the scalars $\phi$ and $\bar\phi$.  These scalar
function coefficients are suppressed below since they will play no
part in our proof, but they should be considered to be present in
any term.  The traces in (\ref{2f}) mean contraction on spinor
indices.  There are other such terms, but they can always be
written as one of the above using a Fierz identity.  Each of these
terms has a definite $\U(1)_R$ charge in $N=2$ harmonic
superspace, and only terms with the same charge can have
variations that cancel.

The first term above, $(\psi^\pm)^2 \;tr(F^2)$, must have either a
$D^+_\a$ or $D^-_\a$ variation that gives an additional $F$,
resulting in a 3-$F$ variation term. This term does not vanish,
since the result is not a Lorentz scalar and does not have the
$F$'s traced.  Since we have no 3-$F$ action terms, and we cannot
get two $F$'s from a single variation, this variation term must be
cancelled by the variation of another 2-$F$ term.  Since there is
only one term with two $F$'s plus fermions for each $\U(1)_R$
charge, there is no other term to cancel this variation.  A
similar argument holds for the second term above, based on the
nonexistence of a term with both an $F$ and an $\bar F$.

To show that the $D^{\pm\pm}\; tr(F^2)$ term cannot appear in the action we
instead look at the term $D^{\pm\pm}\; tr(\bar F^2)$ for simplicity.\footnote{We
do this because we are using the $D^\pm_\a$ variation.  We could just as well
look at the $\bar D^\pm_\ad$ variation of the term $D^{\pm\pm}\; tr(F^2)$, since
$\bar D^{\pm\pm} = D^{\pm\pm}$}  For the terms with $\U(1)_R$ charge 0 or -2
we can act with $D^+_\a$ on the $D^{\pm\pm}$ to get
a term with two $\bar F$'s and a $\bar \psi$:
\be
tr(\bar F^2) \;(\dsl\bar\psi^\pm)_\a .
\ee
(For the term with $\U(1)_R$ charge +2 we instead apply the $D^-_\a$ derivative to
get the same result.) This part of the variation must be cancelled by a 1-$\bar F$
term, since we have no other 2-$\bar F$ terms.  But we cannot get an $\bar F$ field
from a $D^\pm_\a$ variation, and thus there is no way to cancel this variation.
So there can be no terms with two $F$'s or two $\bar F$'s in the action.

\paragraph{Terms with one $F$:}
Now we have to examine terms with one $F$, looking first once again at those
with fermions $\psi^\pm$, $\bar\psi^\pm$.  The possible terms are
\be
(\psi F\psi) \psi^2, \qquad
(\psi F\psi) \bar\psi^2, \qquad
(\psi F\psi) D^{\pm\pm}, \qquad
\psi F\dsl \bar\psi ,
\ee
as well as their conjugates.
The $\pm$ indices on the $\psi$'s, which would indicate the total $\U(1)_R$
charge of each term, have not been included.  It is not hard to show that there
is only one term of each form above for each $\U(1)_R$ charge.  The different
orderings of the $\pm$ indices does not give new terms, as each ordering can
be related to any other using a Fierz identity.

As before, these terms will always have some part of their variation with one more
$F$, giving a 2-$F$ term.  Since we have shown that there are no 2-$F$ terms in
the action, we must cancel this part of the variation with the variation of another 1-$F$
term.  In order for two 1-$F$ terms to give the same 2-$F$ variation, they must have
the same fields, and so in this case must be the same term.  So, the coefficients
of all these 1-$F$ action terms with fermions must be zero.

Here we note that this argument does not hold if there is more than one vector
multiplet.  In that case, there are multiple terms with the same basic field
content.  For example, look at the terms $\psi_1^- F_2 \psi_1^+$ and
$\psi_2^- F_1 \psi_1^+$, where the 1,2 subscripts denote different vector
multiplet ``flavors".
These both give a variation that can be written as $tr(F_1 F_2) \,\psi_{1\a}^+$,
and thus their supersymmetry variation could possibly cancel.  This makes it
much more difficult to determine if there are 3-derivative terms
with more than one vector multiplet.

Now we move on to single-$F$ terms without fermions, which is a little trickier
to explore.  The only possible terms are of the form
\be\label{1f}
(\dsl F)\cdot(\dsl \phi),\qquad
(\dsl F)\cdot(\dsl \bar\phi),\qquad
(\dsl \phi)\cdot F \cdot(\dsl \bar\phi),
\ee
as well as their conjugates.
The two $\dsl$'s in (\ref{1f}) are contracted on dotted indices.  The two derivatives
must act on different fields, because otherwise they would vanish due to the
symmetry of $F$.

Each of these terms requires either $\phi$ or $\bar\phi$.  First we look at the
terms with a $\bar\phi$.  We write these as
\be
(\bar\sigma^\mu F \sigma^\nu)\; \bar\phi\;
f(\phi,\bar\phi)\; \del_\mu\del_\nu,
\ee
where the derivatives must act on different fields.
Consider now $\bar D^+$ acting on the
$\bar\phi$, or, equivalently, the $D^+$ variation of its conjugate.
The conjugate term,
\be
(\sigma^\mu \bar F \bar\sigma^\nu)\; \phi\;
f(\bar\phi,\phi)\; \del_\mu\del_\nu,
\ee
includes in its $D^+$ variation the term
\be\label{1fvar1}
(\sigma^\mu \bar F \bar\sigma^\nu)\; \psi^+_\a\;
f(\bar\phi,\phi)\; \del_\mu\del_\nu.
\ee
We cannot generate an $\bar F$ or a $\phi$ from a $D^+$ variation, and the
only way to get a $\psi^+$ is from the term we are considering.  So, we can only
cancel (\ref{1fvar1}) by generating a $\bar\phi$ from the variation of another
term.  The variation of $\bar\psi^-$ gives $\del\bar\phi$, but this would require a
term in the action with an $F$ and fermions, which we have already show not to
exist.   Thus there can be no term in the action with no fermions, an $\bar F$, and a
$\phi$, and so also no term with no fermions, an $F$, and a $\bar\phi$.

%THIS PART IS NOT NEEDED, BUT I KEEP IT FOR NOW
%But first we must perform a Fierz transformation on the variation term in order
%to put it into a form where the free index $\a$ is on the $\sigma$ contracted
%with the partial derivative acting on $\bar\phi$.  Such a Fierz transforation
%gives us two terms:
%\be
%(\sigma^\nu_{\a\bd} \,(\bar F \bar\sigma^\mu\psi^+)^\bd - \sigma^\mu_{\a\bd}
%\,(\bar F \bar\sigma^\nu\psi^+)^\bd ) \;\; f(\bar\phi,\phi)\; \del_\mu\del_\nu,
%\ee
%Now, we can cancel the first of these terms if the partial derivative $\del_\nu$
%acts on $\bar\phi$, and we can cancel the second if $\del_\mu$ acts on
%$\bar\phi$.  If both the partials act on $\bar\phi$, however, then this term
%vanishes by symmetry, as must the original action term.

So, the only possible 1-$F$, no fermion term left is
\be
(\dsl F)\cdot(\dsl \phi)
\ee
times a function of $\phi$ only, not $\bar\phi$.  This has a variation that includes
\be\label{dFdphivar}
(\dsl F)\cdot(\dsl \psi^+_\a) .
\ee
To get a cancelling variation term we must generate the $F$ field from a $\psi^-$,
\be
(\dsl \psi^+)\cdot(\dsl\psi^-).
\ee
This has a variation which includes
\be
(\dsl \psi^+)\cdot(\dsl F)_\a.
\ee
While this looks similar to (\ref{dFdphivar}), in fact, after performing
Fierz transformations we find that they differ by a term
\be
\dsl_{\a\bd}\psi^{+\ga} \dsl^{\bd \b}F_{\b\ga}.
\ee
There is no way to generate such a term from a $D^+$ variation, and any
Fierz transformation leave us with terms of the form we are trying to cancel
in the first place.  So, there can be no action terms of this form, and so no
action terms with any $F$ fields at all.

%THE FOLLOWING IS THE START OF ANOTHER WAY TO PROVE THIS
%HOPEFULLY IT WILL NOT BE NEEDED, BUT I AM KEEPING IT FOR NOW.
%The variation of these terms all gives a part where the transformation operator
%acts on the $F$.  This gives, after some arrangement, a term of the form
%\be\label{dFdvar}
%f(\phi,\bar\phi)\; (\dsl \bar\psi^+)\cdot(\sigma^{\mu\nu})_\a\; \del_\mu \del_\nu,
%\ee
%where the derivatives $\del_\mu$ and $\del_\nu$ act on two different fields,
%depending on the specific initial field.

%In order to cancel terms of this form, we must generate either $\bar\psi^+$
%or $\dsl \bar\phi$ from a variation, since these are the only fields in the terms
%we can get from a $D^+_\a$ variation.  We can get this only from a $D^{+-}$
%term with two derivatives, such as:
%\be
%(\dsl D^{+-})\cdot(\dsl \phi)
%\ee
%Unlike the term with the $F$, these derivatives must be symmetrized,
%not antisymmetrized, and can be written as:
%\be
%(\del^\mu D^{+-})(\del_\mu \phi)
%\ee
%with variation
%\be
%(\del^\mu (\bar\psi^+))(\del_\mu \phi)
%\ee
%this looks similar in form to \ref{dFdvar}, but in fact they are not the same.
%The term in\ref {dFdvar} has the derivatives antisymmetrized, whereas
%here they are symmetrized.  Thus these terms cannot cancel one another.

%This leaves us with the possibility of generating $\dsl \bar\phi$ from the
%variation of $\bar\psi^-_\ad$.

\paragraph{Terms without $F$:}
Showing that there are no terms without $F$s but with fermions is very similar to
the case with one $F$ and fermions, but even simpler.  Acting on a fermion in
such a term with the appropriate supersymmetry variation once again gives
an $F$, and since we have no 1-$F$ action terms we need to cancel this
variation with the variation of another action term without an $F$.  So, the
$F$ in the variation must be generated in the variation.  Since there is only
one way to get an $F$ from a given variation, it must come from the same
fermion it did before, and so we can only cancel this variation with the same
term we generated it from.  So, the coefficient of a term with fermions and
no $F$'s is zero.

Now all that remains is to show that there are no terms with zero
$F$'s and no fermions.  The only fields we have left are the
$D$'s, the $\phi$'s, and their derivatives.  Without $\psi$'s the
derivatives must contract with each other, and so we must have two
or zero derivatives.  This means we must have one or three $D$'s,
respectively.  The variation of a three-$D$ term cannot be
cancelled by a one-$D$ term, and it is easy to see that distinct
3-$D$ terms cannot cancel, so we can have no three-$D$ terms.  We
can ignore the $\U(1)_R$ charges of the $D$'s now, since only
terms with the same net charge can possibly cancel, and so the
charges on the $D$'s will just go along for the ride.  The only
remaining possible terms are then \be D \partial^2 \phi
f(\phi,\bar\phi),\qquad D \partial^2 \bar\phi
g(\phi,\bar\phi),\qquad D \partial \phi \partial \bar\phi
h(\phi,\bar\phi). \ee The variation of these terms must then
cancel among themselves. A $D^+$ variation acting on the scalar
functions will give a $\psi$ for each term.  These three terms
then each have derivatives on different fields, and we cannot move
the derivatives around using integration by parts because of the
$D$.  This means that $f$, $g$, and $h$ must be independent of
$\phi$, as well as $\bar\phi$, and therefore just constants.  Now
if we let the variation act on the $D$ field, we get three terms
with $\dsl\psi$ parts, and the other parts will not cancel, and do
not form a total derivative.  So there are no terms that have
neither $F$'s nor $\psi$'s.

This exhausts all possible terms, and shows that we cannot have any
supersymmetric, gauge invariant, three-derivative terms with
only one vector multiplet.  For reasons discussed above, it seems much
more difficult to generalize this argument to theories with more than one vector
multiplet.  Similarly, this strategy becomes quite cumbersome to use to
search for Chern-Simons-like terms at the four-derivative level.

\section{Global issues on the Coulomb branch}

The conclusion of the negative arguments of the last two
sections is that the issue of the existence of Chern-Simons-like
terms in particular, and of a systematic and effective derivative
expansion in general, in $N=2$ effective actions on the Coulomb
branch is problematic.  In the $N=1$ and $N=0$ cases
similar combinatoric problems connected with gauge
invariance and the existence of Chern-Simons-like terms arise.
But in these cases because the gauge fields can be assigned
positive derivative weight, this does not present a problem
of principle for the derivative expansion.  For $N\ge2$ theories,
however, the gauge potential superfield must be assigned
negative weight in the derivative expansion, and the existence
of Chern-Simons-like terms becomes a problem of principle for
the existence of a systematic derivative expansion.

One positive result of the searches in sections 3 and 4 was the
identification of a 4-derivative superspace Chern-Simons-like term
globally on the Coulomb branch of $N=2$ theories, which could not 
have been found had we worked solely with field strength superfields.  
(Note that such a term might also survive on the moduli space
of $N=4$ theories, provided that it can be completed
to an $N=4$ supersymmetric multiplet \cite{bi0111}.)
We will devote the remainder of this section to a discussion of this term
and the related issue of global obstructions on the Coulomb branch.
This issue has interesting parallels to the recent discussion in
\cite{bw0409} of F terms on
the moduli space of $N=1$ theories which are globally obstructed
from being written as D-terms.

Recall from section 3 that the term in question is the four-derivative
term
\be\label{glob1}
S_4 = \int\!d^4xdud^8\th\,
V_a^{--}\, \Dp \left(A^b_a(W,\Wb) \Dp W_b\right) + \mbox{c.c.},
\ee
where $A$ satisfies
\be\label{glob2}
\del^c A^b_a - \del^b A^c_a = 0 ,
\ee
which is the local integrability condition for 
\be\label{glob3}
A^b_a = \del^b B_a,
\ee
for some $B_a(W,\Wb)$, where $\del^b \equiv \del/\del W_b$
is the holomorphic derivative on the Coulomb branch, $\cal M$. 
If (\ref{glob3}) held, then 
$S_4$ could be written solely in terms of field strength superfields
as $S_4 \sim \int\!d^4xd^8\th\, B_a(W,\Wb)\, \Wb^a$. But (\ref{glob2})
is only a local integrability condition, so $B_a$ may fail to exist
globally on the Coulomb branch.

Indeed, treating $A_a^b$ as the coefficients functions of a set of
(1,0)-forms on the Coulomb branch,
\be\label{glob4}
A_a \equiv A_a^b(W,\Wb)\, dW_b,
\ee
condition (\ref{glob2}) is equivalent to the $A_a$ being closed under
the Dolbeault exterior differential $\del \equiv dW_a \del^a$,
\be
\del A_a = 0,
\ee
which implies, locally, that the $A_a$ are exact,
\be
A_a = \del B_a ,
\ee
for some (0,0)-forms $B_a$.  So, the interesting Chern-Simons-like
terms are the non-trivial Dolbeault cohomology classes in
$H^{(1,0)}_\del({\cal M})$.  

There are, however, some caveats to this description of the global
Chern-Simons-like terms, which come from the low energy $\U(1)^n$ 
gauge invariance.  First, since the potential superfield appears
explicitly in (\ref{glob1}), the coefficient functions $A_a^b$, and
therefore the one-forms $A_a$, are only defined up to holomorphic 
linear redefinitions of the $W_a$, instead of general holomorphic
changes of variables on $\cal M$.  This is because the gauge 
invariance of (\ref{glob1}) under $\de V^{--}_a = -\Dmm\la_a$
does not permit non-linear transformations of the $V^{--}_a$, and
the $W_a$ are linearly related to the $V^{--}_a$ by (\ref{Wdef}).
We are thus restricted to special coordinates on the Coulomb 
branch in (\ref{glob1}).

Secondly, the special coordinates, $W_a$, on $\cal M$ are not single-valued.
They are allowed, by virtue of the electric-magnetic duality ambiguity in the 
description of $\U(1)^n$ theories, to have monodromies valued in the discrete
$\Sp(2n,\Z)$ duality group \cite{sw9407}.  These monodromies are determined 
by the 2-derivative terms in the effective action on the Coulomb branch and 
transform the field strength superfields, $W_a$, nonlinearly and the potential 
superfields, $V^{--}_a$, in a nonlocal way.  This makes the global definition of 
the Chern-Simons-like term problematic.  It would be desirable to have a 
duality-covariant superspace formalism for the vector multiplets in order to 
address this issue.  For the field strength multiplet, it is not hard to develop 
such a formalism along the lines of \cite{ss9304}. For the potential superfield,
needed in the Chern-Simons-like term (\ref{glob1}), such a formalism is not known.

However, there are some models in which the electric-magnetic monodromies
are trivial and so (\ref{glob1}) can be used.  The simplest
example is the Coulomb branch of the scale invariant $\SU(2)$ $N=2$
superQCD with four massless fundamental hypermultiplets.  In this
case the Coulomb branch is $\C^*$, the complex $W$-plane minus
the point at the origin, and the special coordinate $W$ experiences
only the $\Z_2$ monodromy $W\to-W$ upon circling the origin (inherited from
the un-gauge-fixed center of the $\SU(2)$ gauge group).  Thus
the coefficient function $A$ in (\ref{glob1}) is constrained to be $\Z_2$-even:
$A=A(W^2, \Wb^2,W\Wb)$.  Any such $AdW$ is trivially closed under $\del$;
unfortunately, it is also exact.  (For example, $AdW=(\Wb/W)dW=
\del[\Wb\ln(W\Wb)]$.)
Other monodromy-free Coulomb branches occur for scale-invariant
theories with $\SU(2)^n$ product gauge groups with fundamental and
bi-fundamental matter in various configurations \cite{w9703}; but all
of the resulting non-compact Coulomb branches appear to have trivial
$H^{(1,0)}_\del$.

A less familiar set of theories which do have non-trivial cohomology
classes are the
$N=2$ theories with compact Coulomb branches discussed in 
\cite{i9708,gk9802,i9909}.  These theories can be realized as
compactifications of 6-dimensional little string theories \cite{brs9704,s9705}
on $T^2$.  In the simplest example \cite{gk9802}, the Coulomb branch is given
in special coordinates by the $\Z_2$ orbifold of a complex torus minus the
four orbifold fixed points; \ie, the complex $W$-plane with the
identifications $W\sim -W$, $W\sim W+1$, and $W\sim W+\Lambda$,
for some complex scale $\Lambda$, minus the four points 
$W\in\{0,{1\over2},{\Lambda\over2},{1+\Lambda\over2}\}$.
(This is the scale invariant $\SU(2)$ model of the last paragraph
with toroidally compactified Coulomb branch.)  A nontrivial element
of $H^{(1,0)}_\del$ has a constant coefficient function $A$, so that
the (1,0)-form is $\sim dW$ which is not exact since $W$ is not single-valued on
$\cal M$, nor can any function of $\Wb$ be added to it to make it 
single-valued.\footnote{More 
generally, since $T^2/\Z_2$ minus its fixed points is equivalent to
a four-punctured sphere, there is a rich set of analytic functions $g$ on
this space, and $A = g(\Wb) dW$ is non-trivial in cohomology; presumably
physical considerations will place limits on the allowed singularities in $g$
at the fixed points.}  This thus gives an example of a global Chern-Simons-like
term.  Inserting this $A$ in (\ref{glob1}) shows that the existence of this
term simply stems from the fact that although $W$ is not single-valued on
$T^2$, $\Dp W$ is.

The global structure of the Coulomb branch also plays a role in the
classification of the holomorphic 4-derivative term listed in (\ref{Wterms}),
which has the form
\bea\label{glob6} 
S_{4a} &=& \int\! d^4x\,
d^4\th \,\, \del_\mu W_a \del^\mu W_b \, {\cal G}^{ab}(W) +\mbox{c.c.} ,
\nonumber\\
&=& {1\over2} \int\! d^4x\, d^4\th\, (\Dbp)^2
[\Dm W_a \cdot \Dm W_b \, {\cal G}^{ab}]  +\mbox{c.c.},
\eea
where in the second line we have used the $N=2$ algebra
(\ref{Dalg}) and the chirality of $W$ (\ref{CC}).
This shows that this term is more properly thought of
as an integral over 3/4 of superspace; it is a special case 
of one of the 3/4 superspace terms found in \cite{aabe0306}.
Because of its holomorphic nature, terms of this form enjoy a
non-renormalization theorem, but only so long as they cannot be
rewritten as a nonholomorphic term integrated over the full
superspace.
The following manipulations give a condition for when this
can happen.
\bea\label{glob7}
 \int\!\! d^4x\, d^4\th\, d^2\thbp \Biggl\{
 \Dm W_a \cdot \Dm W_b \, K^{ab}
&=& -W_a \, (\Dm)^2 W_b \, K^{ab}
- W_a \Dm W_b \cdot \Dm W_c \, \del^c K^{ab}
\nonumber\\
&=& -W_a \, (\Dbm)^2 \Wb_b \, K^{ab}
- \Dm W_a \cdot \Dm W_b \, W_c \del^b K^{ac}\Biggr\},
\nonumber
\eea
where in the second line we used the Bianchi identity (\ref{bianchi}).
Moving the last term to the left side then gives
\be\label{glob8}
\int\!\! d^4x\, d^4\th\, d^2\thbp \,  \Dm W_a \cdot \Dm W_b \,  \del^b[W_cK^{ac}]
= -8 \int\!\! d^4x\, d^8\th\, W_a  \Wb_b \, K^{ab}.
\ee
Thus, if the ${\cal G}$ holomorphic coefficient function in (\ref{glob6}) 
satisfies
\be\label{glob9}
{\cal G}^{ab}(W) = \del^b H^a(W)
\ee
for some holomorphic $H^a = W_c K^{ac}$, then the $S_{4a}$ term can be 
rewritten as a non-holomorphic 4-derivative term integrated over the whole 
superspace.  With two or more vector multiplets, (\ref{glob9}) can fail to be 
integrable even locally, giving examples of locally holomorphic $S_{4a}$ terms.

But, even with just a single vector multiplet, (\ref{glob9}) can fail to be integrable 
globally.  In this case of a single vector multiplet, the globally holomorphic 
coefficient function ${\cal G}$ in (\ref{glob6}) can be thought of as defining 
a section of a holomorphic quadratic form ${\cal G}(W)\,(dW)^2$ on the 
Coulomb branch, $\cal M$.  By (\ref{glob9}) this global coefficient function is 
defined up to the equivalence
\be
{\cal G}(W) \sim {\cal G}(W) + \del H(W),
\ee
which, because of the holomorphy of the functions, is the same as the equivalence
in holomorphic de Rham cohomology on the Coulomb branch (as opposed to
Dolbeault cohomology).  As in our discussion of the Chern-Simons-like term 
above, when there are non-trivial electric-magnetic duality monodromies on 
the Coulomb branch the global definition of the ${\cal G}$ section is more 
complicated, and cannot be taken simply as single-valued on $\cal M$.  It is
an interesting open question whether the holomorphic 4-derivative terms
(\ref{glob6}) exist by virtue of a global obstruction on one-dimensional Coulomb
branches.

\acknowledgments
It is a pleasure to thank C. Beasley, M. Edalati, O. Lunin, R. Rabadan, A. Shapere, 
E. Witten, and P. Yi for helpful comments and discussions.
This work was supported in part by DOE grant DOE-FG02-84ER-40153.
P.C.A. was supported in part by a grant in aid from the IBM Einstein
Endowed Fellowship.
G.A.B. was supported in part by a University of Cincinnati URC grant.

\end{document}